\documentclass[preprint,aps,showkeys,showpacs]{revtex4}
\usepackage{graphicx}

\begin{document}

\title{A Fast 3D Poisson Solver with Longitudinal Periodic and Transverse 
Open Boundary Conditions for Space-Charge Simulations}

\author{Ji Qiang}
\affiliation{
Lawrence Berkeley National Laboratory,
          Berkeley, CA 94720}

\begin{abstract}

A three-dimensional (3D) Poisson solver with longitudinal periodic and transverse
open boundary conditions can have important applications in
beam physics of particle accelerators.
In this paper, we present a fast efficient method to solve
the Poisson equation using a spectral finite-difference method.
This method uses a computational domain that contains the charged
particle beam only and
has a computational complexity of $O(N_u(logN_{mode}))$,
where $N_u$ is the total number of unknowns and $N_{mode}$ is the maximum
number of longitudinal or azimuthal modes.
This saves both the computational time and the memory usage by using an
artificial boundary condition in a large extended computational domain.

\end{abstract}

\pacs{52.65.Rr; 52.75.Di}
\keywords{Poisson solver, spectral finite-difference method,
periodic and open boundary conditions}

\maketitle

\section{Introduction}

The particle accelerator as one of the most important inventions of the twenty
century has many applications in science and industry.
In accelerators, a train of charged particle
(e.g. proton or electron) beam bunches are transported and accelerated
to high energy for different applications.
To study the dynamics of those charged particles self-consistently
inside the accelerator, 
the particle-in-cell (PIC) model is usually employed in 
simulation codes (e.g. the WARP and the IMPACT code suite~\cite{friedman,qiang1,impact-t}).
This PIC model includes both the space-charge
forces from the Coulomb interactions among the charged particles
within the bunch and the forces from external accelerating and focusing fields
at each time step.
To calculate the space-charge forces, one needs to solve
the Poisson equation for a given charge density distribution.
A key issue in the PIC simulation
is to solve the Poisson equation efficiently, at each time step, subject
to appropriate boundary conditions. 

Solving the 3D Poisson equation
for the electric potential of a charged beam bunch 
with longitudinal periodic and transverse open boundary conditions
can have important applications in beam dynamics study of 
particle accelerators. 
In the accelerator, a train of charged particle bunches as shown
in Fig. 1 are produced,
accelerated, and transported. If the separation between two bunches is
large, each bunch can be treated as an isolated bunch,
and the 3D open boundary conditions can be used to solve the Poisson equation.
In some accelerators such as a Radio-Frequency Quadrupole (RFQ),
the separation between particle bunches is short, to model a single bunch,
one needs to use
the longitudinal periodic boundary condition~\cite{rfq}. 
The same model can be used to study
space-charge effects in a longitudinally modulated electron beam,
where the electron beam density varies periodically from the interaction
with the laser beam and the magnetic optic elements~\cite{dohlus}. 
\begin{figure*}
\includegraphics[angle=0,width=.60\textwidth]{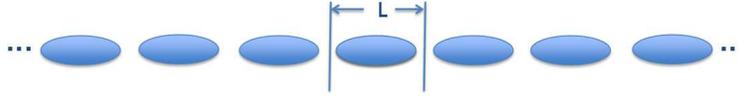}
\caption{A schematic plot of a train of charged particle beam bunches in the particle accelerator.}
\end{figure*}


In previous studies, 
a number of methods for solving 3D Poisson's equation subject to a variety
of boundary conditions
have been 
studied~\cite{haidvogel,ohring,hockney,dang,braverman,plagne,qiang2,qiang3,lai,peter,xu,rob,mads,zheng,qiang4,anderson}.
However, none of these methods handles the Poisson equation with the
longitudinal periodic and transverse open boundary conditions.
In the code of reference~\cite{qiang1}, an image charge method is used to add the contributions from
longitudinally periodic bunches into the single bunch's Green function.
Then an FFT method is used to effectively calculate the discrete convolution
between the charge density and the new Green's function that includes 
contributions from other bunches. The computational cost of this
method scales as $O(Nlog(N))$. However, this method requires the 
computation of the Green's function from multiple bunch summation. It is
not clear, how many bunches are needed in order to accurately
emulate the longitudinal periodic boundary condition.
In reference~\cite{dohlus}, the image charge method is used with
special function to approximate
the summation of the Green's function in different regimes. 
In practical application, one may
not know beforehand what regime should be used for a good approximation.
%
Besides the complexity of the new Green's function in the image charge
method,
to use the FFT to calculate the discrete convolution, one
needs to double the computational domain with zero padding~\cite{hockney,nr}.
This increases both the computational time and the memory usage.

In this paper, we propose a fast efficient method
to solve the 3D Poisson equation with the longitudinal
periodic and transverse open boundary conditions. We  
use a Galerkin spectral Fourier method to approximate the electric potential
and the charge density function in the longitudinal and azimuthal dimensions where
periodic boundary conditions are satisfied.
We then use a second order finite-difference method to solve the 
radial ordinary differential equation for each mode subject to
the transverse open boundary condition. Instead of using a large
radial domain with empty space and artificial finite Dirichlet 
boundary condition to approximate
the open boundary condition, we use a domain that contains only the
charged particle beam and a boundary matching condition to 
close the group of 
linear algebraic equations for each mode. 
This group of tridiagonal linear algebraic
equations can be solved efficiently using the direct Gaussian 
elimination with a computational cost $O(N)$, where
$N$ is the number of unknowns on the radial grid.

The organization of this paper is as follows:
After the introduction, we describe the proposed spectral 
finite-difference numerical method
in Section~II. 
Several numerical tests of the 3D Poisson solver 
are presented in Section~III. 
The conclusions are drawn in
Section~IV.

\section{Numerical Methods}

%

The three dimensional Poisson equation in
cylindric coordinates can be written as:
\begin{equation}
	\label{3dpoi}
\frac{\partial^2 \phi}{\partial r^2} +
\frac{1}{r}\frac{\partial \phi}{\partial r} + \frac{1}{r^2}
\frac{\partial^2 \phi}{\partial \theta^2} +
\frac{\partial^2 \phi}{\partial z^2} = - \rho(r,\theta,z)
\end{equation}
where $\phi$ denotes the electric potential, 
$\rho$ the charge
density function, $r$ and $z$ the radial 
and longitudinal distance. 
The longitudinal periodic and transverse open 
boundary conditions for the potential are:
\begin{eqnarray}
\phi(r=\infty,\theta,z) & = & 0  \\
\phi(r,\theta + 2\pi,z) & = & \phi(r,\theta,z)  \\
	\phi(r,\theta,z+L) & = &  \phi(r,\theta,z)
\end{eqnarray}

Given the periodic boundary conditions of the electric potential along
the $\theta$ and the $z$, we use a Galerkin spectral method with the Fourier 
basis function to approximate the charge density function $\rho$ and the electric potential $\phi$ along these two dimensions as:
\begin{eqnarray}
	\rho(r,\theta,z) & = & \sum_{n=-N_n/2}^{n=N_n/2-1} 
	\sum_{m=-N_m/2}^{m=N_m/2-1} \rho_n^m(r) \exp(-i a_nz)\exp(-im \theta) \\
	\phi(r,\theta,z) & = & \sum_{n=-N_n/2}^{n=N_n/2-1} 
	\sum_{m=-N_m/2}^{m=N_m/2-1} \phi_n^m(r) \exp(-i a_nz)\exp(-im \theta) 
\end{eqnarray}
where 
\begin{eqnarray}
	\rho_n^{m}(r) &  = & \frac{2}{L\pi}\int_0^{L}\int_0^{2\pi} \rho(r,\theta,z) \exp(im\theta)\exp(i a_n z) \ d\theta dz \\
	\phi_n^{m}(r) & = & \frac{2}{L\pi}\int_0^{L}\int_0^{2\pi} \phi(r,\theta,z) \exp(im\theta)\exp(i a_n z) \ d\theta dz 
\end{eqnarray}
and $a_n = n 2\pi/L$, $L$ is the longitudinal periodic length.
Substituting the above expansions into the Poisson Eq.~\ref{3dpoi} and making use
of the orthonormal condition of the Fourier function, we obtain:
\begin{eqnarray}
	\frac{\partial^2 \phi^m_n}{\partial r^2} + \frac{1}{r}
	\frac{\partial \phi_n^m}{\partial r} -(\frac{m^2}{r^2}+(a_n)^2)
	\phi_n^m & = & -{\rho_n^m}
	\label{rmn}
\end{eqnarray}
This is a group of decoupled ordinary differential equations that can
be solved for each individual mode $m$ and $n$. For these equations,
at $r=0$, we have the boundary conditions:
\begin{eqnarray}
	\frac{\partial \phi_n^m }{\partial r} (0) & = & 0 ; \ \ for \ \ m = 0 \\
	\phi_n^m (0) & = & 0 ; \ \ for \ \ m \neq 0
\end{eqnarray}
Assuming all charged particles within the beam bunch are contained within a
radius $R$, we discretize the above equation 
using a second order finite-difference scheme,
and obtain a group of linear algebraic equations for each mode $(m,n)$ as:
\begin{eqnarray}
	(\frac{r_i^2}{h^2} - \frac{r_i}{2 h}) \phi_n^m(r_{i-1}) -
	(\frac{2 r_i^2}{h^2}+m^2+a_n^2 r_i^2) \phi_n^m(r_i) +
	(\frac{r_i^2}{h^2} + \frac{r_i}{2 h}) \phi_n^m(r_{i+1}) = 
	-r_i^2 {\rho_n^m(r_i)}
\end{eqnarray}
where $i=1,2,\cdots,N$, and $r_i = i h$. The boundary conditions
at $r=0$ are approximated as:
\begin{eqnarray}
	-\frac{3}{2} \phi_n^m (r_0) +2\phi_n^m(r_1) -\frac{1}{2} \phi_n^m(r_2) & = & 0 ; \ \ for \ \ m = 0 \\
	\phi_n^m (r_0) & = & 0 ; \ \ for \ \ m \neq 0
\end{eqnarray}
For $m=0$, there are only $N+1$ linear equations but $N+2$ unknowns, and 
for $m\neq 0$, there are only $N$ linear equations but $N+1$ unknowns.
For the potential outside the radius $R$, the Eq.~\ref{rmn} can be
written as:
\begin{eqnarray}
	\frac{\partial^2 \phi^m_n}{\partial r^2} + \frac{1}{r}
	\frac{\partial \phi_n^m}{\partial r} -(\frac{m^2}{r^2}+(a_n)^2)
	\phi_n^m & = & 0
	\label{rmn0}
\end{eqnarray}
subject to the open boundary conditions
\begin{eqnarray}
	\phi_n^m(r=\infty) & = & 0 
	\label{r0bc}
\end{eqnarray}
For $n\neq 0$, a formal solution of the equation \ref{rmn0} subject to the boundary condition
\ref{r0bc} can be written as:
\begin{eqnarray}
	\phi_n^m(r) & = & A \ K_m(a_n r)
	\label{phi0}
\end{eqnarray}
where $K_m$ is the second kind modified Bessel function. 
Using the above equation 
and the continuity of the potential at $r_N$,
we obtain another equation for the unknowns $\phi_n^m(r_N)$ and $\phi_n^m(r_{N+1})$ as:
\begin{eqnarray}
\phi_n^m(r_N) K_m(a_n r_{N+1}) & = & \phi_n^m(r_{N+1}) K_m(a_n r_{N})
	\label{bc1}
\end{eqnarray}
For $n=0$,$m\neq0$, the Eq.~\ref{rmn0} is reduced to the
Cauchy-Euler equation:
\begin{eqnarray}
	\frac{\partial^2 \phi^m_0}{\partial r^2} + \frac{1}{r}
	\frac{\partial \phi_0^m}{\partial r} -\frac{m^2}{r^2}
	\phi_0^m & = & 0
	\label{rmn02}
\end{eqnarray}
A formal solution of this equation that satisfies the open radial boundary condition
can be written as:
\begin{eqnarray}
	\phi_0^m(r) & = & A r^{-m}
	\label{phi02}
\end{eqnarray}
From the above equation, 
we obtain another equation for the unknowns $\phi_0^m(r_N)$ and $\phi_0^m(r_{N+1})$ as:
\begin{eqnarray}
	\phi_0^m(r_N) r_N^{m} & = & \phi_0^m(r_{N+1}) r_{N+1}^{m}
	\label{bc2}
\end{eqnarray}
For $n=0$,$m=0$, the Eq.~\ref{rmn0} is reduced to:
\begin{eqnarray}
	\frac{\partial^2 \phi^0_0}{\partial r^2} + \frac{1}{r}
	\frac{\partial \phi_0^0}{\partial r} 
	& = & 0
	\label{rmn03}
\end{eqnarray}
A formal solution of this equation that satisfies the open radial boundary
condition can be written as:
\begin{eqnarray}
	\phi_0^0(r) & = & A \log(r)
	\label{phi03}
\end{eqnarray}
From this equation, 
we obtain another equation for the unknowns $\phi_0^0(r_N)$ and $\phi_0^0(r_{N+1})$ as:
\begin{eqnarray}
	\phi_0^0(r_N) \log(r_{N+1}) & = & \phi_0^0(r_{N+1}) \log(r_N)
	\label{bc3}
\end{eqnarray}
Using Eqs.~\ref{bc1}, \ref{bc2}, \ref{bc3}, 
we have $N+2$ linear equations for $N+2$ unknowns for $m=0$ and 
$N+1$ linear equations for $N+1$ unknowns for $m\neq0$.
For each mode $m$ and $n$, this is a group of tridiagonal linear algebraic 
equations, which can be solved
effectively using direct Gaussian elimination with the number of operations
scaling as $O(N)$. 
Since both Fourier expansions in $\theta$ and $z$ can be computed
very effectively using the FFT method, the total computational complexity
of the proposed algorithm scales as $O(N N_m N_n log(N_mN_n))$.

\section{Numerical Tests}

The numerical algorithm discussed in the preceding section is
tested using two charge density distribution functions.
The first example is an infinite long cylindric coasting beam with uniform
charge distribution within the radius $R = 2$.
The charge density function is given as
\begin{eqnarray}
\rho(r,\theta,z) & = & 
\left\{ \begin{array}{r@{\quad:\quad}l}
1.0 & r \le 2  \\
                0.0              & r > 2
                        \end{array} \right.
\end{eqnarray}
For this charge density function, there is only the radial component of the 
electric
field. The analytical solution of the electric field can be
found as:
\begin{eqnarray}
	E_r (r) & = & \frac{r}{2} \ \ for \ r\le 2
\end{eqnarray}

\begin{figure*}
\includegraphics[angle=270,width=.45\textwidth]{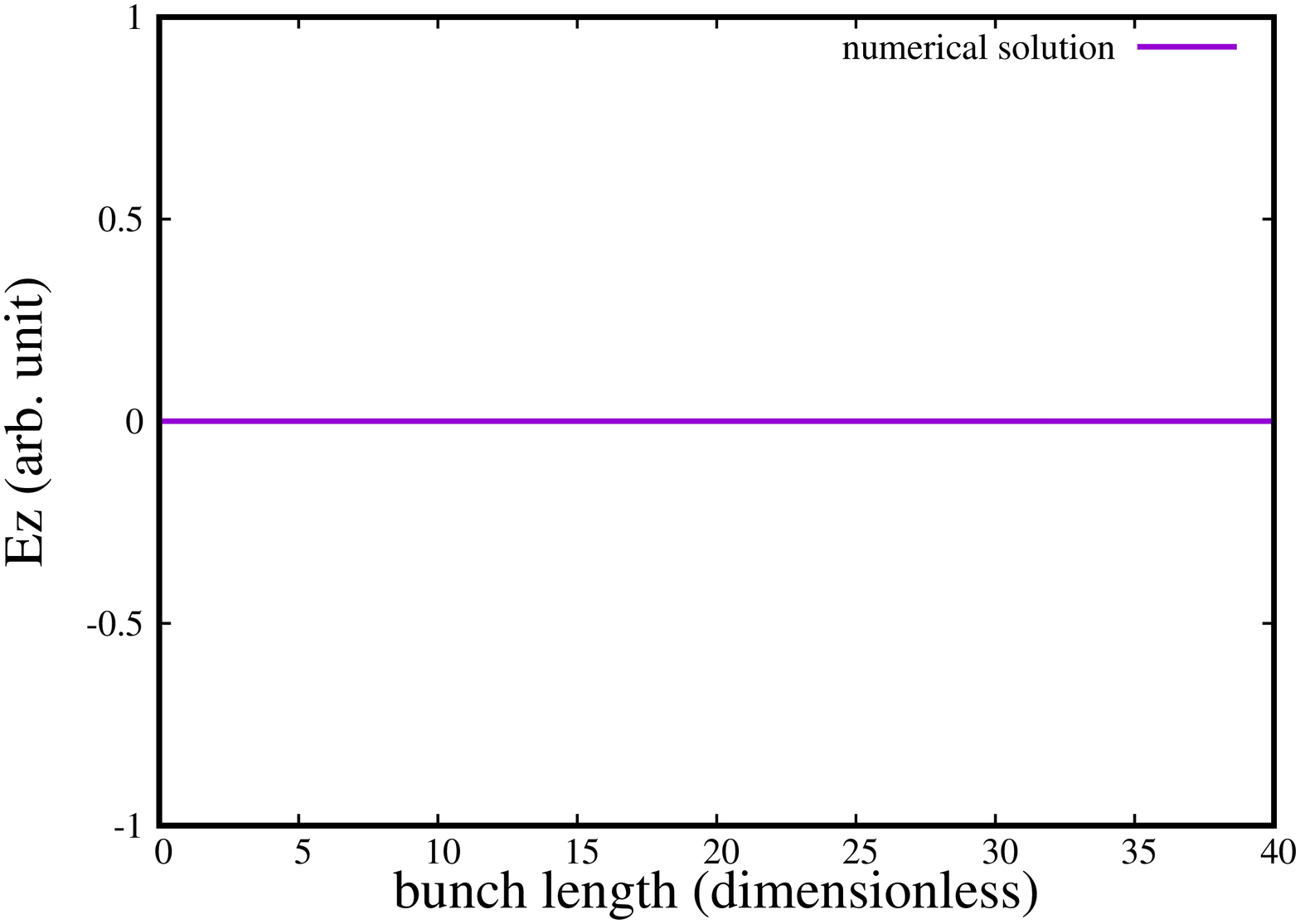}
\includegraphics[angle=270,width=.45\textwidth]{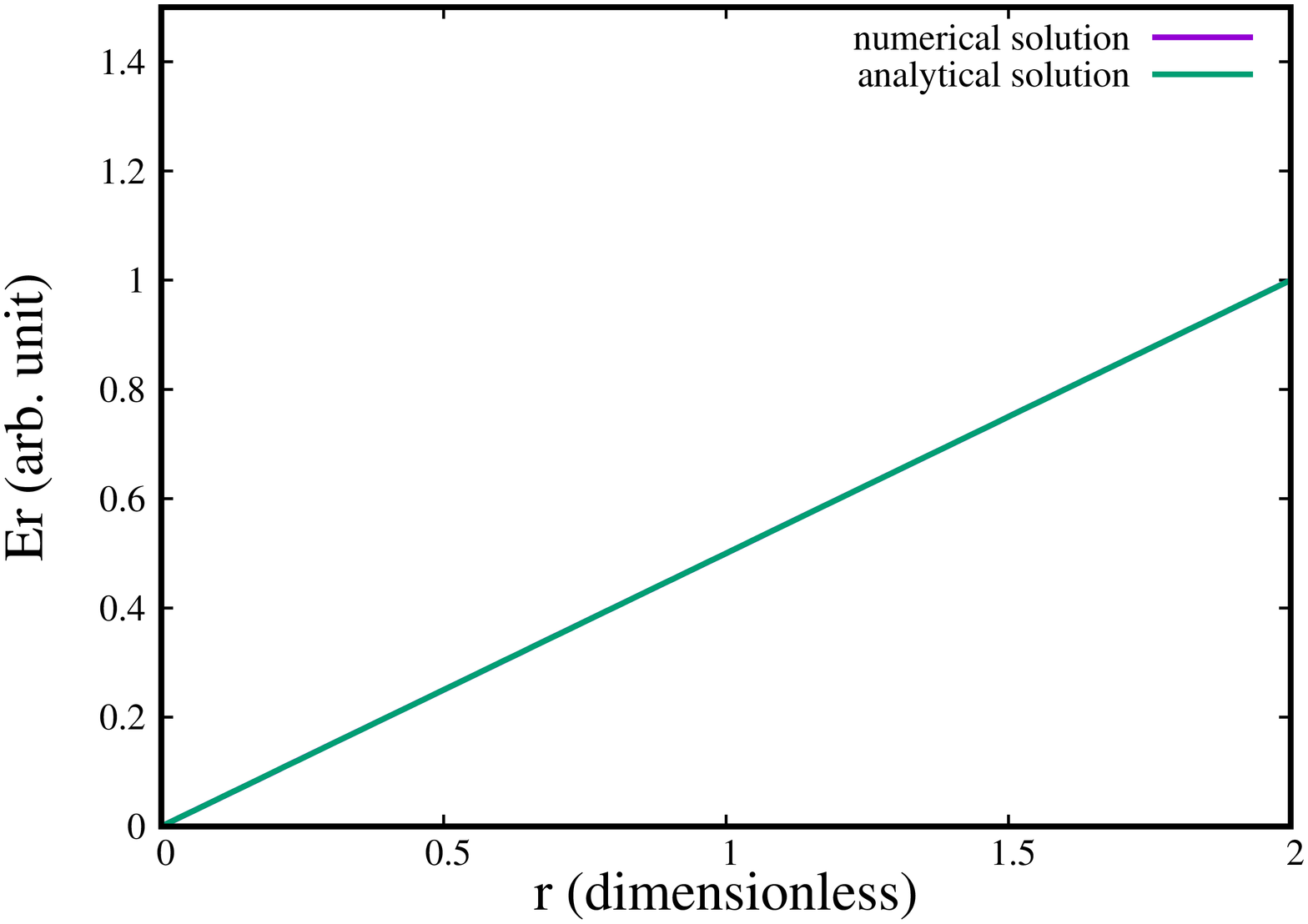}
\caption{Longitudinal electric field profile Ez on the z-axis (left) and the 
	transverse radial
	electric field profile Er in the middle of the bunch 
	from the numerical solutions 
	together with the analytical solution (right) in a uniform
	cylinder coasting beam.}
	\label{fig2}
\end{figure*}

Figure~\ref{fig2} shows the longitudinal electric field and the transverse 
radial electric field from the numerical solution and the above analytical
solution. It is seen that the numerical solution agrees with the analytical
solution very well.

\begin{figure*}
\includegraphics[angle=270,width=.45\textwidth]{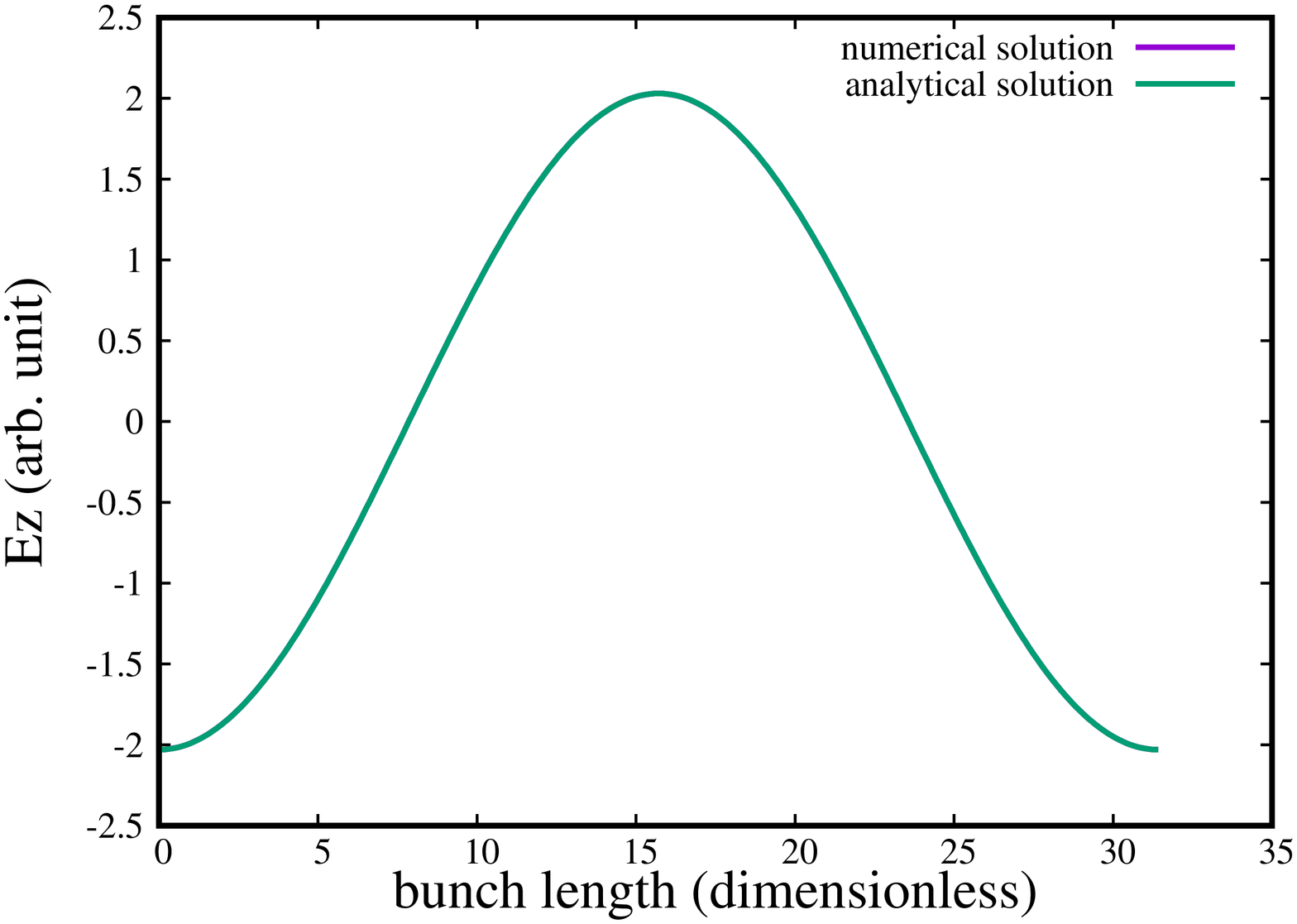}
\includegraphics[angle=270,width=.45\textwidth]{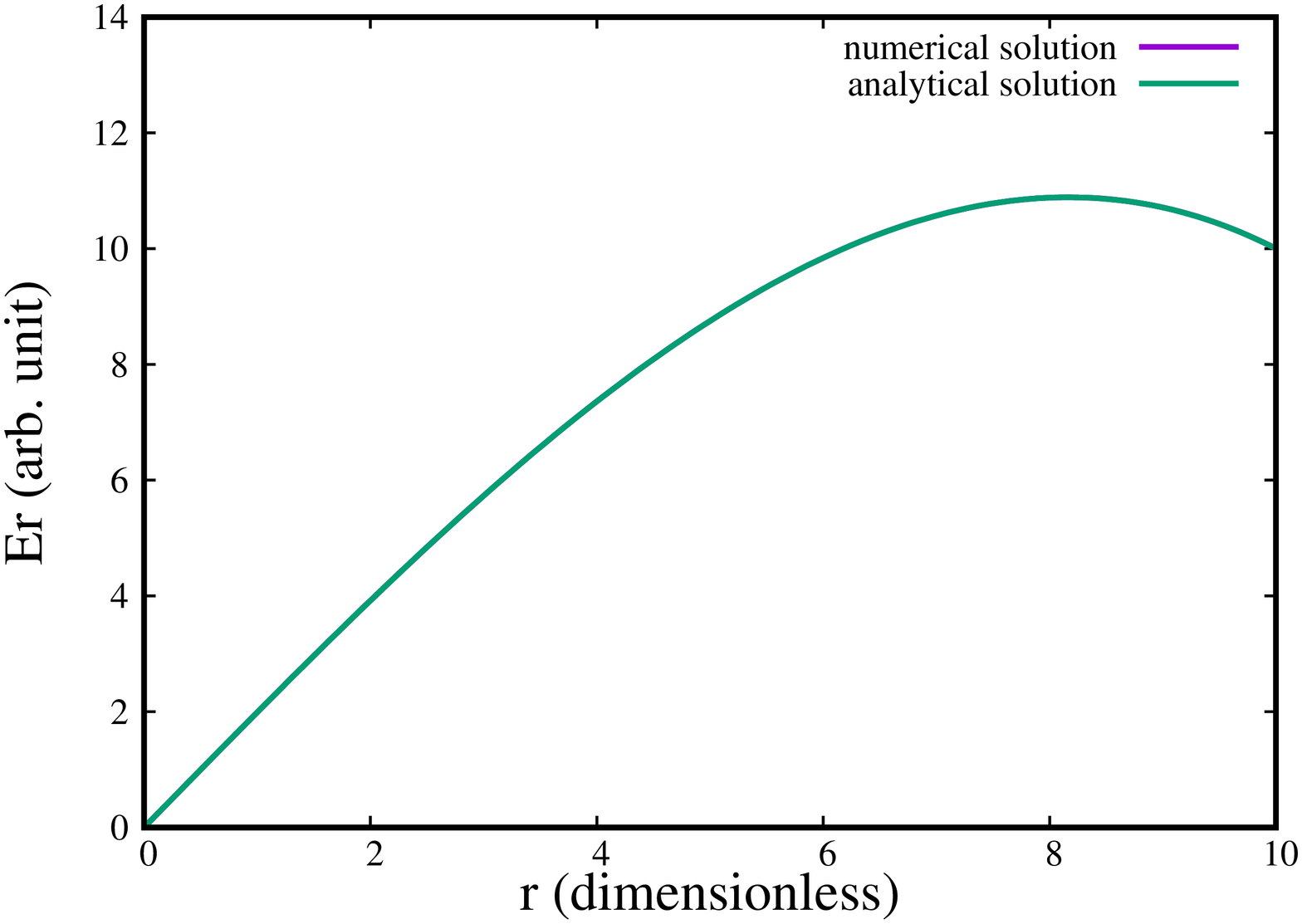}
\caption{Longitudinal electric field profile Ez on the z-axis (left) and the 
	transverse radial
	electric field profile Er in the middle of the bunch 
	(right) from the numerical solutions 
	together with the analytical solutions in a 
	longitudinally modulated charged particle beam bunch.}
	\label{fig3}
\end{figure*}

In the second test example, we assume that there is a longitudinal modulation
of the charged particle density distribution. The charge density function
is given as:
\begin{eqnarray}
\rho(r,\theta,z) & = & 
\left\{ \begin{array}{r@{\quad:\quad}l}
		4-4(r/R)^2+\sin(a_1z)[4-(a_1 r)^2]/5 & r \le R  \\
                0.0              & r > R
                        \end{array} \right.
\end{eqnarray}
The analytical solution of the electric fields for this charge distribution 
can be written as:
\begin{eqnarray}
	E_z(r,z) & = & a_1 \cos(a_1 z)[r^2 - A I_0(a_1 r)]/5  \\
	E_r(r,z) & = & 2r -r^3/R^2 +\sin(a_1 z)[2r-A a_1 I_1(a_1 r)]/5
\end{eqnarray}
where the constant $A$ is given as:
\begin{eqnarray}
	A & = & \frac{R^2a_1 K_1(a_1 R) +2RK_0(a_1R)}{a_1 I_1(a_1 R) K_0(a_1 R)
+ a_1 I_0(a_1 R) K_1(a_1 R) }
\label{aa}
\end{eqnarray}
Here, the matching condition at the edge $R$
is used together with the analytical formal 
solution Eq.~\ref{phi0} for the open boundary condition to determine
the above constant $A$.
Figure~\ref{fig3} shows the longitudinal electric field and the transverse
radial electrical field from the numerical solutions and from the
analytical solutions. The numerical solutions and
the analytical solutions agree
with each other very well in this longitudinally modulated charged particle
beam too. Here, we have assumed $R=10$ and $L=\pi R$. 

The numerical method proposed in the preceding section
has the advantage that uses a computational domain that contains the 
charged particle beam only while satisfying the transverse open
boundary condition. In principle, the transverse open boundary can
be approximated by an artificial closed Dirichlet boundary condition
in a larger computational domain. Since only the electric fields inside
the charge particle beam bunch are needed in the self-consistent accelerator
space-charge beam dynamics simulation, this larger computational domain by using the
artificial Dirichlet boundary condition will waste both the computational
time and the memory storage in the empty computational domain.
In the following, we use a simplified one-dimensional equation from
above equations to illustrate the advantage of the above proposed method.

\begin{figure*}
\includegraphics[angle=270,width=.45\textwidth]{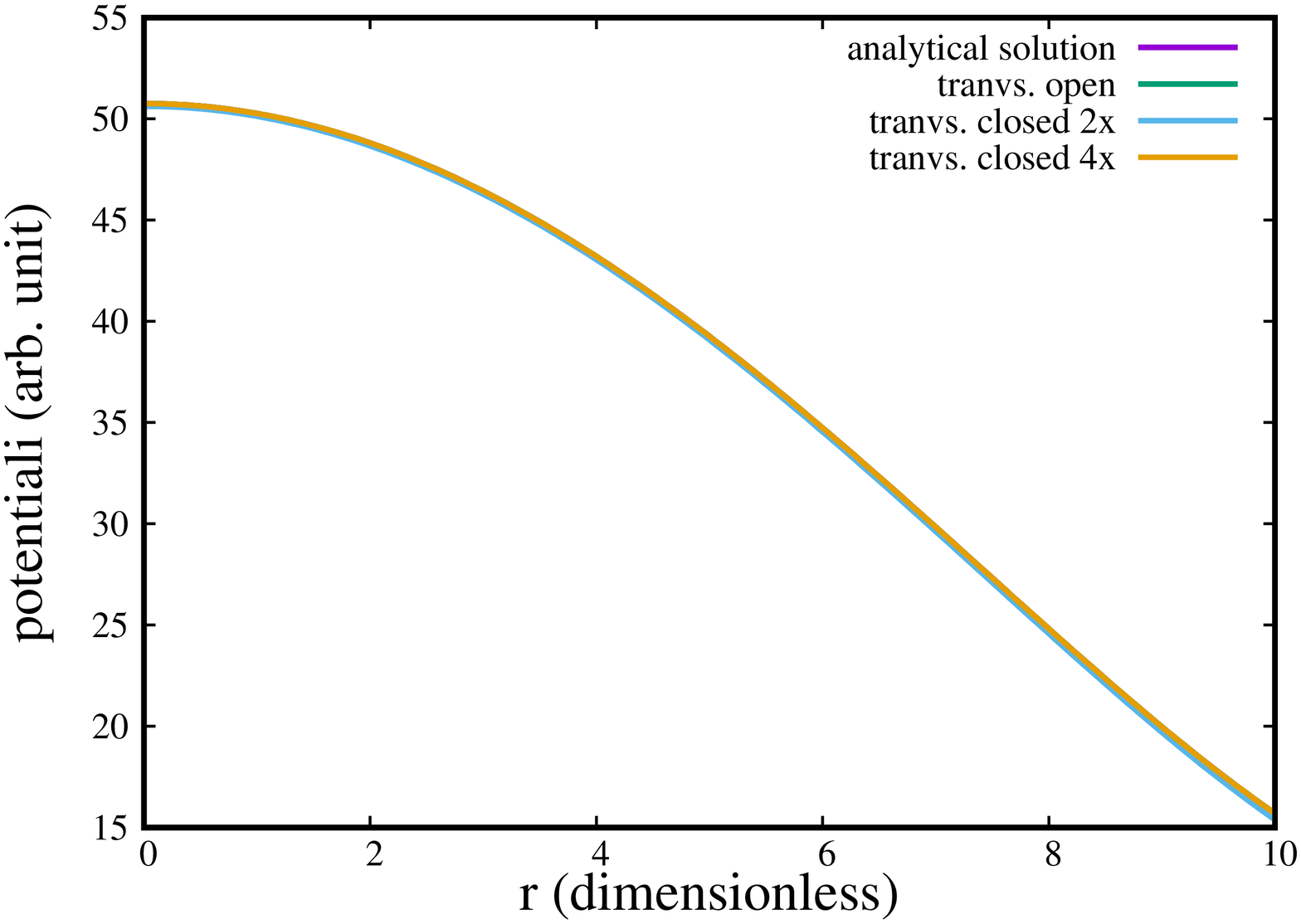}
\includegraphics[angle=270,width=.45\textwidth]{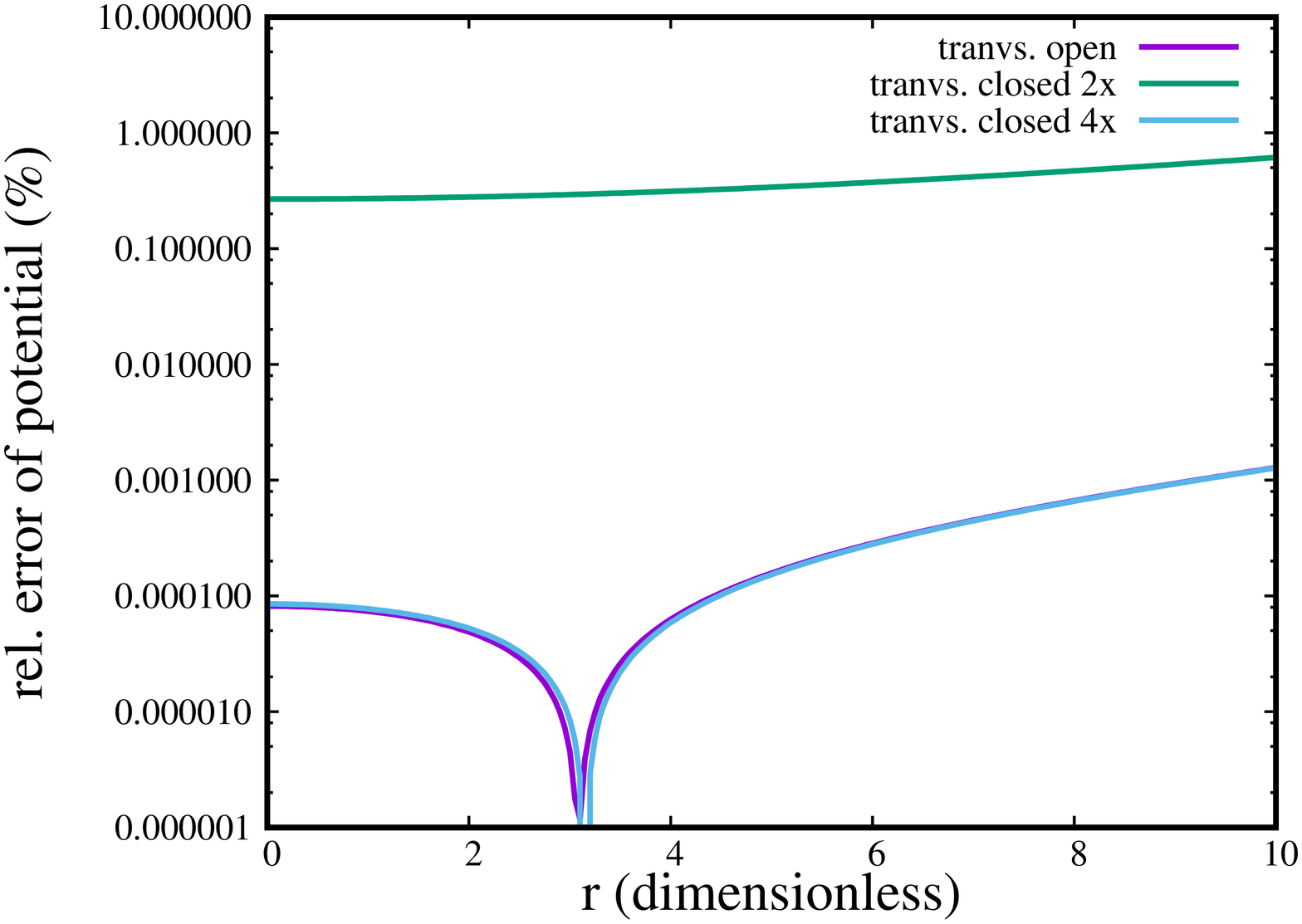}
\caption{The 
	electric potential (left) and the relative errors (right) 
	from the analytical solution and from
	the numerical solution with transverse
	open, transverse closed with two times radial computational domain,
        and transverse closed with four times computational domain.}
	\label{fig4}
\end{figure*}

For $m=0$ and $n=1$, Eq.~\ref{rmn} is reduced
to:
\begin{eqnarray}
	\frac{\partial^2 \phi^0_1}{\partial r^2} + \frac{1}{r}
	\frac{\partial \phi_1^0}{\partial r} - a_1^2
	\phi_1^0 & = & -\rho_1^0
	\label{rmn00}
\end{eqnarray}
Assuming a radial charge distribution $\rho_1^0(r)$ as:
\begin{eqnarray}
	\rho_1^0(r) & = & 
\left\{ \begin{array}{r@{\quad:\quad}l}
		4-(a_1 r)^2 & r \le R  \\
                0.0              & r > R
                        \end{array} \right.
\end{eqnarray}
we can have an analytical solution as:
\begin{eqnarray}
	\phi^0_1(r) & = & -r^2 + A I_0(a_1 r)
\end{eqnarray}
where the constant $A$ is given in Eq.~\ref{aa}.
Figure~\ref{fig4} shows the electric potential and the relative errors 
as a function of radial
distance from the analytical solution, and from the proposed
numerical solution with transverse open boundary condition ($R=10$),
from the artificial transverse closed Dirichlet boundary
condition using two times computational domain ($\phi_1^0(R=20) = 0$),
and from the artificial transverse closed boundary condition
using four times computational domain ($\phi_1^0(R=40) = 0$). 
It is seen that even using two times computational domain, the 
artificial closed boundary condition solution still shows much larger
errors than the proposed open boundary numerical solution.
It appears that four times larger computational domain is needed
in the artificial closed boundary solution in order to attain 
the same numerical accuracy as the open boundary solution that uses
a domain with radius $R$ that contains the charged particle beam only.
In the above example, we have used $201$ radial grid points
for the open boundary solution, $401$ grid points for the artificial closed
boundary solution with two times computational domain and $801$ grid
points for the solution with four times computational domain.

\section{Conclusions}

In this paper, we presented a fast three-dimensional Poisson solver
subject to longitudinal periodic and transverse open boundary conditions.
Instead of using a larger artificial computational domain with closed
Dirichlet boundary condition, this solver uses a computational domain
that contains the charged particles only. This saves both the computational
time and the memory usage compared with the artificial closed boundary
condition method. By using the FFT method to calculate the longitudinal
and azimuthal Fourier expansion and the direct Gaussian elimination to
solve the radial tridiagonal linear algebraic equations, the computational
complexity of the proposed numerical method scales as $O(N_u(logN_{mode}))$.
This makes this fast Poisson solver very efficient and can be included
in the self-consistent space-charge simulation PIC codes for
space-charge beam physics study in particle accelerators.

\section{ACKNOWLEDGEMENTS}
This work was supported by the U.S. Department of Energy under Contract 
No. DE-AC02-05CH11231.
This research used computer resources at the National Energy Research
Scientific Computing Center.

\end{document}